\documentclass{main}

\usepackage[english]{babel}
\usepackage{cite}

\usepackage{gensymb}

\usepackage{placeins}

\usepackage{epsf}
\usepackage{amssymb}
\usepackage{amsmath}
\DeclareMathOperator{\Tr}{Tr}
\usepackage{amsfonts}
\usepackage{psfrag,epsfig,graphics}
\usepackage{soul}

\usepackage{bm,slashed,multirow,
	soul,mathtools,xspace,array,comment}                
\allowdisplaybreaks
\usepackage{bbold}
\usepackage{xcolor}
\usepackage{subfigure}
\usepackage[printonlyused]{acronym}
\usepackage{hyperref}
\usepackage{tikz}
\usepackage[compat=1.1.0]{tikz-feynman}

\tikzset{cross/.style={cross out, draw=black, minimum size=2*(#1-\pgflinewidth), inner sep=0pt, outer sep=0pt},
	cross/.default={1pt}}
\tikzset{
	on each segment/.style={
		decorate,
		decoration={
			show path construction,
			moveto code={},
			lineto code={
				\path [#1]
				(\tikzinputsegmentfirst) -- (\tikzinputsegmentlast);
			},
			curveto code={
				\path [#1] (\tikzinputsegmentfirst)
				.. controls
				(\tikzinputsegmentsupporta) and (\tikzinputsegmentsupportb)
				..
				(\tikzinputsegmentlast);
			},
			closepath code={
				\path [#1]
				(\tikzinputsegmentfirst) -- (\tikzinputsegmentlast);
			},
		},
	},
	mid arrow/.style={postaction={decorate,decoration={
				markings,
				mark=at position .5 with {\arrow[#1]{stealth}}
	}}},
}

\usepackage{graphicx}

\usepackage{siunitx}
\usepackage{booktabs}

\usepackage{float}

\usepackage{textcomp}
\usepackage{fancyhdr}
\usepackage{lastpage}
\def\be{\begin{equation}}
\def\ee{\end{equation}}
\def\bea{\begin{eqnarray}}
\def\eea{\end{eqnarray}}

\newcommand{\FIG}{Fig.~}

\newcommand{\SEC}{Sec.~}

\newcommand{\EQ}{Eq.~}
\newcommand{\EQs}{Eqs.~}

\fancypagestyle{firstpagefooter}{%
  \fancyfoot[L]{  \textcopyright \footnotesize This work is an open access article under a Creative Commons Attribution 4.0 International License (https://creativecommons.org/licenses/by/4.0/).}
  \fancyfoot[C]{}  
   
}

\begin{document}

\thispagestyle{firstpagefooter}
\title{\Large Exclusive photoproduction of a $\pi^0\gamma $ pair in the saturation framework }

\author{$^{1,2}$M. Fucilla, $^{1,3}$S. Nabeebaccus, $^{2}$L. Szymanowski,  $^{1}$ S. Wallon, $^{1}$\underline{J. Yarwick}\footnote{Speaker, email: joseph.yarwick@ijclab.in2p3.fr}}

\address{
$^{1}$Université Paris-Saclay, CNRS/IN2P3, ĲCLab, 91405 Orsay, France
 \\ $^{2}$National center for Nuclear Research (NCBJ), 02-093 Warsaw, Poland
  \\
$^{3}$Department of Physics \& Astronomy, University of Manchester, Manchester M13 9PL, United Kingdom
}

\maketitle
\abstracts{
We consider the exclusive photoproduction of a $\pi^0 \gamma$ pair with large invariant mass, as a promising channel to study the effects of gluon saturation. It has recently been demonstrated that this process is incompatible with a collinear factorization approach in terms of generalized parton distributions (GPDs) at the leading twist. In such a situation, a (generalized) $k_T$-dependent factorization at small $x$ is a valid alternative approach. We perform this calculation using the shockwave formalism, which resums multiple gluon exchanges between the projectile and the dense nuclear target. We find that the polarized amplitude changes sign as a function of back-to-back transverse momentum $|\vec{p}_t|$ of the pion-photon pair, resulting in a dip-like structure in the fully differential cross section as a function of $|\vec{p}_t|$.
}

\footnotesize DOI: \url{}

\keywords{QCD, shockwave, saturation, exclusive processes, small $x$}

\section{Introduction}

		 In ultraperipheral collisions (UPCs), strong interactions between the incoming hadrons are suppressed, and thus on-shell photon exchanges become dominant. For perturbative QCD (pQCD) to be applicable, one necessarily requires a hard scale. For the process under study, namely the exclusive photoproduction of a $\pi^0\gamma$ pair,
        \begin{align}
            \gamma(p_\gamma) + N(p_N) \longrightarrow \gamma(k) + N'(p_{N'}) + \pi^0(p_{\pi})\,,
        \end{align}
        this is provided by the large invariant mass of the photon-pion pair in the final state, $M_{\pi^0\gamma}^2$.
        
        The exclusive photoproduction of  a $\pi^0 \gamma$ pair is a rather peculiar process, since it has been shown in \cite{Nabeebaccus:2023rzr,Nabeebaccus:2024mia} that it breaks collinear factorization, due to the presence of a Glauber pinch. This is linked to the fact that this channel allows for 2-gluon exchanges with the nucleon sector. Indeed, in cases where they are absent \cite{Boussarie:2016qop,Duplancic:2018bum,Duplancic:2022ffo,Duplancic:2023kwe}, on the basis of electric charge and charge parity conservation, a collinear factorization approach is indeed valid \cite{Qiu:2022bpq,Qiu:2022pla}.

        Here, we instead adopt a $k_T$ factorization approach at small $x$.  To take gluon saturation into account, we work within the shockwave formalism \cite{McLerran:1994vd, Balitsky:1995ub}. This formalism has been the framework for numerous recent studies of diffractive processes \cite{Boussarie:2014lxa, Boussarie:2016ogo, Boussarie:2016bkq, Boussarie:2019ero, Fucilla:2022wcg, Fucilla:2023mkl, Boussarie:2024pax, Boussarie:2024bdo}. In this framework, $t$-channel gluons are separated by a rapidity cutoff and distinguished by what are referred to as internal gluons, those with forward momenta larger than the cutoff and are absorbed into the projectile impact factor, and external gluons, those which are eikonally approximated to have zero forward momentum upon boosting the reference frame to that of the dilute projectile. This then admits an effective Lagrangian which may be used to sum all diagrams of increasing number of external gluon exchanges to produce effective Feynman rules in terms of a shockwave background represented by a Wilson line exchange. In the case of a diffractive color singlet exchange, the color trace of Wilson lines is expressed in terms of a dipole operator which naturally evolves in $x$  according to the B-JIMWLK equations. Furthermore, in such a $k_T$ factorization approach, the end-point singularities that occur in collinear factorization calculation are naturally regularized by the transverse momenta of the $t$-channel gluons.
        
\section{Computation}
		\label{sec:Computation}

        \subsection{Kinematics}
        \label{subsec:kinematics}
		As the shockwave approach is defined in the high energy limit, it is natural to work in lightcone coordinates. We define our lightcone basis as
		\begin{gather}
        \label{eq:lightcone-vectors}
        n_1^{\mu} \equiv \frac{1}{\sqrt{2}}(1, 0,0, 1)\,, \qquad n_2^{\mu} \equiv \frac{1}{\sqrt{2}}(1, 0, 0, -1).
        \end{gather}
        Any 4-momentum can then be decomposed as
        \begin{align}
        \label{eq:4-momentum}
            p^{\mu} = p^+n_1^{\mu}+p^-n_2^{\mu}+p_{\perp}^{\mu},
        \end{align}
        such that
		\begin{align}
			&p^+ = p \cdot n_2 = \frac{1}{\sqrt{2}}(p^0+p^3),\qquad p^-=p\cdot n_1 = \frac{1}{\sqrt{2}}(p^0-p^3) ,\nonumber
			\\ 
            & p\cdot q = p^+q^-+p^-q^++p_{\perp}\cdot q_{\perp} = p^+q^-+p^-q^+-\vec{p}\cdot \vec{q}.
		\end{align}
        We work in the limit of semi-hard kinematics,
		\begin{equation}
        \label{eq:semi-hard-kinematics}
			s_{\gamma N}\gg M^2_{\pi^0\gamma}\gg\Lambda_{\mathrm{QCD}}^2\,.
		\end{equation}
		$s_{\gamma N}$ is the center-of-mass energy of the incoming photon-target system, where we omit the subscript in future reference for convenience. $M^2_{\pi^0\gamma}$ is the invariant mass squared of the pion and outgoing photon which we take to be the hard scale in the photoproduction limit. The projectile momentum is approximated to be fully in the $n_1$ ($+$) direction while the target momentum is oriented in the $n_2$ ($-$) direction. This implies that 
		\begin{equation}
			p_\gamma^+, \; p_N^- \sim \sqrt{\frac{s}{2}}\,,\qquad \mathrm{with}\qquad 
			p_{\gamma}^+ \sim p_q^+ \sim p_{\Bar{q}}^+ \gg 
			p_N^+\,,\quad p_N^- \gg p_{\gamma}^-, \; p_q^-, \; p_{\Bar{q}}^-\,,
		\end{equation}
        where $p_q$ and $p_{\bar q}$ are the momenta of the quark and anti-quark that will eventually form the pion, $p_{\pi} = p_{q}+p_{\bar q}$.
		In addition, we work in the frame such that the incoming photon has zero transverse momentum.  The momenta can then be parameterized as
		\begin{align}
			\label{eq:momenta}
            	p_{\gamma}^{\mu} &= p^+_{\gamma}n_1^{\mu} -\frac{Q^2}{2p_{\gamma}^+}n_2^{\mu}\,,\qquad p_N^{\mu} = \frac{s}{2p_{\gamma}^+}n_2^{\mu}\,,
            \qquad p_{N^\prime}^{\mu} = p_N^{\mu} + \Delta_{\perp}^{\mu}\,,\\
			p_{\pi}^{\mu} &= \alpha_{\pi}p_{\gamma}^+n_1^{\mu}+\frac{(\vec{p}_t+\frac{\vec{\Delta}}{2})^2}{2p_{\gamma}^+\alpha_{\pi}}n_2^{\mu}-p_{\perp}^{\mu}-\frac{\Delta^{\mu}_{\perp}}{2},\qquad
			k^{\mu} = \alpha_{k}p_{\gamma}^+n_1^{\mu}+\frac{(\vec{p}_t-\frac{\vec{\Delta}}{2})^2}{2p_{\gamma}^+\alpha_{k}}n_2^{\mu}+p_{\perp}^{\mu}-\frac{\Delta^{\mu}_{\perp}}{2}\,,
		\end{align}
		where $\Delta_{\perp}^{\mu}$ is the total $t$-channel momentum transferred to the target and $\vec{p}_t$ is the back-to-back momentum of the pion-photon in the transverse plane. By conservation of momentum,
		\begin{equation}
			\alpha_\pi+\alpha_k=1\,.
		\end{equation}
        Note that the virtuality of the incoming photon is kept general here, though we will later assume the photoproduction limit, $Q^2 = 0$. In such a case, the relevant transverse photon polarization vectors are
        \begin{align}        \label{eq:polarization-vectors}\epsilon^\mu_T(p_\gamma)=\epsilon_{\gamma\perp}^\mu\,\;\;,\qquad
		      \epsilon^{\mu}_{T}(k) = \left(\frac{\vec{\epsilon}_{k\perp} \cdot \vec{k}}{k^+}\right) n_2^{\mu} +\epsilon_{k\perp}^{\mu}\,,
		\end{align}
        where we have used the lightcone gauge $A\cdot n_2 = 0$. 
        The perpendicular component of the polarization vectors can be defined by a linear polarization configuration taking the $x$-axis to be along $\vec{p}_t$,
        \begin{align}
        \label{eq:linear-polar}
				 \epsilon^{j}_{\perp x} &= 
                 \frac{p^j_{\perp}}{|\vec{p}_t|}\,,
				\qquad
				 \epsilon^{j}_{\perp y} = 
                 -\frac{\varepsilon^{+-p_{\perp}j}}{|\vec{p}_t|}\,,
			\end{align}
    where $\varepsilon^{+-p_{\perp}j}$ corresponds to the partially contracted Levi-Civita tensor, $n_{1\mu} n_{2\nu} p_{\perp\alpha} \varepsilon^{\mu \nu \alpha j}$, and $j$ is a transverse index.

    \begin{figure}
			\centering
			\begin{center}
				\begin{tikzpicture}[scale=0.45]
					\draw[decorate,decoration=snake,draw=black] (-4,-1)--(0,-1);
					\path[draw=black,postaction={on each segment={mid arrow=black}}] (1.5,0.5) arc (90:180:1.5);
					\path[draw=black,postaction={on each segment={mid arrow=black}}] (0,-1) arc (180:270:1.5);
					\path[draw=black,postaction={on each segment={mid arrow=black}}] (5,0.5) -- (1.5,0.5);
					\path[draw=black,postaction={on each segment={mid arrow=black}}] (1.5,-2.5) -- (5,-2.5);
					\path[double,line width=1.8pt, draw=black] (-4,-5) -- (5,-5);
					\filldraw[fill=gray, fill opacity=1](2,-2.5) ellipse (0.3cm and 4cm);
					\draw[->] (-3, -0.7) -- (-1, -0.7);
					\draw[->] (3.5, -2.2) -- (4.5, -2.2);
					\draw[->] (3.5, 0.8) -- (4.5, 0.8);
					\draw[->] (1.4, -3.7) -- (1.4, -2.7);
					\draw[->] (1.4, -0.7) -- (1.4, 0.3);
					\node at (4, -1.9) {\textcolor{black}{\footnotesize $p_q$}};
					\node at (4, 1.1) {\textcolor{black}{\footnotesize $p_{\bar{q}}$}};
					\node at (1, -3.2) {\textcolor{black}{\footnotesize $\Vec{p}_1$}};
					\node at (1, -0.2) {\textcolor{black}{\footnotesize $\Vec{p}_2$}};
					\node at (-2, -0.4) {\textcolor{black}{\footnotesize $p_{\gamma}$}};
					\node at (-4.4,-1) {\textcolor{black}{\footnotesize $\gamma$}};
					\node at (5.2, -2.5) {\textcolor{black}{\footnotesize $q$}};
					\node at (5.2, 0.5) {\textcolor{black}{\footnotesize $\Bar{q}$}};
					\node at (-5, -5) {\textcolor{black}{\footnotesize $N(p_N)$}};
					\node at (6.1, -5) {\textcolor{black}{\footnotesize $N^{\prime}(p_N^{\prime})$}};
					
					\draw[decorate,decoration=snake,draw=black] (2.8,0.5)--(5,3);
					\draw[->] (3.2, 1.5) -- (4, 2.5);
					\node at (3.5, 2.4) {\textcolor{black}{\footnotesize $k$}};
					
					\filldraw[black] (2.8,0.5) circle (2pt);
					\filldraw[black] (0,-1) circle (2pt);
					\node at (0.4, -1) {\textcolor{black}{\footnotesize $y_0$}};
					\node at (2.7, 0.2) {\textcolor{black}{\footnotesize $y_1$}};
					\node at (-6.5, -1.5) {\textcolor{black} \Large {$M_{q\bar{q},1}=$}};


					
				\end{tikzpicture}
                \qquad
				\begin{tikzpicture}[scale=0.45]
					\draw[decorate,decoration=snake,draw=black] (-4,-1)--(0,-1);
					\path[draw=black,postaction={on each segment={mid arrow=black}}] (1.5,0.5) arc (90:180:1.5);
					\path[draw=black,postaction={on each segment={mid arrow=black}}] (0,-1) arc (180:270:1.5);
					\path[draw=black,postaction={on each segment={mid arrow=black}}] (5,0.5) -- (1.5,0.5);
					\path[draw=black,postaction={on each segment={mid arrow=black}}] (1.5,-2.5) -- (5,-2.5);
					\path[double,line width=1.8pt, draw=black] (-4,-5) -- (5,-5);
					\filldraw[fill=gray, fill opacity=1](2,-2.5) ellipse (0.3cm and 4cm);
					\draw[->] (-3, -0.7) -- (-1, -0.7);
					\draw[->] (3.5, -2.2) -- (4.5, -2.2);
					\draw[->] (3.5, 0.8) -- (4.5, 0.8);
					\draw[->] (2.6, -3.7) -- (2.6, -2.7);
					\draw[->] (2.6, -0.7) -- (2.6, 0.3); 
					\node at (4, -1.9) {\textcolor{black}{\footnotesize $p_q$}};
					\node at (4, 1.1) {\textcolor{black}{\footnotesize $p_{\bar{q}}$}};
					\node at (3, -3.2) {\textcolor{black}{\footnotesize $\vec{p}_1$}};
					\node at (3, -0.2) {\textcolor{black}{\footnotesize $\vec{p}_2$}};
					\node at (-2, -0.4) {\textcolor{black}{\footnotesize $p_{\gamma}$}};
					\node at (-4.4,-1) {\textcolor{black}{\footnotesize $\gamma$}};
					\node at (5.2, -2.5) {\textcolor{black}{\footnotesize $q$}};
					\node at (5.2, 0.5) {\textcolor{black}{\footnotesize $\Bar{q}$}};
					
					\draw[decorate,decoration=snake,draw=black] (1.1,0.4)--(5,3);
					\draw[->] (2.5, 1.8) -- (4, 2.8);
					\node at (3.1, 2.7) {\textcolor{black}{\footnotesize $k$}};
					
					\filldraw[black] (1.1,0.4) circle (2pt);
					\filldraw[black] (0,-1) circle (2pt);
					\node at (0.4, -1) {\textcolor{black}{\footnotesize $y_0$}};
					\node at (1.1, 0) {\textcolor{black}{\footnotesize $y_1$}};
					
					
					\node at (-6.5, -1.5) {\textcolor{black} \Large {$M_{q\bar{q},2}=$}};
					
					\node at (-5, -5) {\textcolor{black}{\footnotesize $N(p_N)$}};
					\node at (6.1, -5) {\textcolor{black}{\footnotesize $N^{\prime}(p_N^{\prime})$}};
					
				\end{tikzpicture}
			
            \end{center}
           \caption{LO amplitude of photon scattering with a dense nuclear target via shockwave interaction, producing a real photon and neutral pion. Two representative diagrams are shown, for the case of photon emitted from the antiquark, after (left) and before (right) passing through the shockwave background. The total momentum transfer to the target $\Delta = -p_1-p_2$.}

        \label{fig:LO-diag-1}
					\end{figure}
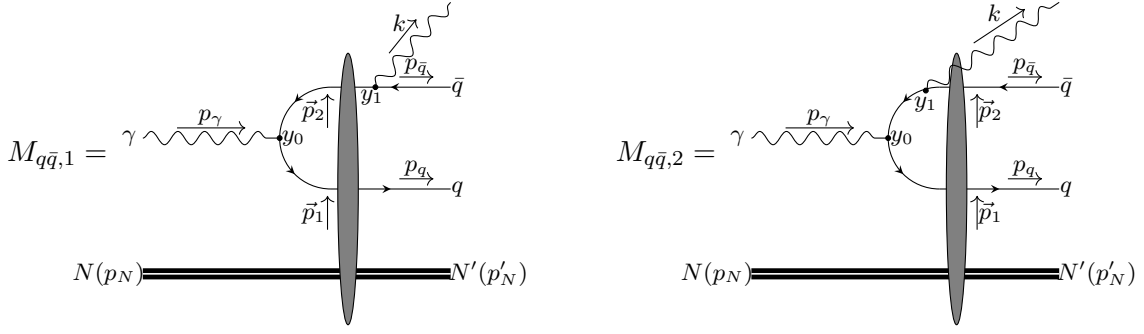

\subsection{Projectile impact factor}

        We first calculate the amplitude $M_{q\bar{q}}$ at the level of $\gamma\rightarrow\gamma q \bar{q}$, amputated of the the initial and final target states, using the shockwave formalism. There are a total of 4 diagrams, 2 of which are shown in Fig. \ref{fig:LO-diag-1}, whereas the other 2 are simply related by symmetry in $q\leftrightarrow\bar{q}$. This symmetry involves taking the Dirac adjoint and reversing the flow of fermionic momentum which results in the exchange of $p_{q}\leftrightarrow p_{\bar{q}}$, $p_1\leftrightarrow p_2$ and $\bar{u}_{p_q} \leftrightarrow v_{p_{\bar{q}}}$ and a reversal of Dirac matrices.

        The full contribution of the amplitude to the S-matrix, including the non-interacting part (indicated by the overline), and with open index $\mu$ corresponding to the polarization of the incoming photon, is
        \begin{align}
			\overline{M}_{q\bar{q},1}^{\mu} = & \int d^4 y_0 d^4 y_1 \theta(-y_0^+) \theta(y_1^+) \frac{\delta_{ij}}{\sqrt{N_c}} \left[\bar{u}(p_q, y_0)\right]^{im}_{0>y_0^+} \left(-ie_q \gamma^{\mu} \frac{e^{-ip_{\gamma}\cdot y_0}}{\sqrt{2p_{\gamma}^+}}\right) \nonumber
			\\
			& \times \left[G(y_0, y_1)\right]^{mj}_{y_1^+>0>y_0^+} \left( -ie_q\hat{\epsilon}_T^*(k)  \frac{\theta (k^+) e^{ik\cdot y_1}}{\sqrt{2k^+}}\right) \left( v_{p_{\bar{q}}} \frac{\theta(p^+_{\bar{q}})e^{ip_{\bar{q}} \cdot y_1}}{\sqrt{2p_{\bar{q}}^+}} \right)\,,
            \\
			\overline{M}_{q\bar{q},2}^{\mu} = & \int d^ 4 y_0 d^4 y_1 \theta(-y_0^+) \theta(-y_1^+) \frac{\delta_{ij}}{\sqrt{N_c}} \left[\bar{u}(p_q, y_0)\right]^{im}_{0>y_0^+} \left(-ie_q \gamma^{\mu} \frac{e^{-ip_{\gamma}\cdot y_0}}{\sqrt{2p_{\gamma}^+}}\right)\nonumber
				\\
				&
				\times G(y_0, y_1) \left( -ie_q\hat{\epsilon}_T^*(k)  \frac{\theta (k^+) e^{ik\cdot y_1}}{\sqrt{2k^+}}\right) \left[ v(p_{\bar{q}}, y_1)\right]^{mj}_{0>y_1^+}, 
		\end{align}
        where the elements in square brackets are the effective shockwave Feynman rules and contain the respective Wilson lines with Latin indices to indicate color matrix components \cite{Boussarie:2016txb}. Note, we have also used the hat notation for contractions with Dirac gamma matrices, i.e. $\hat{p} \equiv \gamma^\nu p_{\nu}$. After carrying out the coordinate space integrals, the interacting part of the amplitude, constructed by subtracting the non-interacting part where the Wilson lines are set to unity, generally reads
        \begin{align}
        \label{eq:quark-antiquark-amp}
            M_{q\bar{q}}^{\mu}& = \overline{M}_{q\bar{q}}^{\mu}-\overline{M}_{q\bar{q}}^{\mu}|_{U,U^{\dagger} \rightarrow\;1} = 
            \\
            \notag & \frac{\delta(p^+_{q}+p^+_{\bar{q}} + k^+ - p_{\gamma}^+)}{\sqrt{2p_\gamma^+}\sqrt{2p_{q}^+}\sqrt{2p_{\bar{q}}^+} \sqrt{2k^+}} \left(\frac{e_q^2\sqrt{N_c}}{(2\pi)}\right)\int d^2 p_{1\perp}d^2 p_{2\perp}\delta^2(p_{q\perp}+p_{\bar{q}\perp}+k_\perp - p_{1\perp}-p_{2\perp})\,\tilde{\mathcal{U}}_{12}\Phi_{q\bar{q}}^{\mu}(p_{1\perp},p_{2\perp}),
        \end{align}
        where we have also pulled out several prefactors from the definition of the impact factor, including $e_q = eQ_q$. Here, $\tilde{\mathcal{U}}_{12}$ is the dipole operator in momentum space defined by 
        \begin{equation}
            \tilde{\mathcal{U}}_{12} = \int d^2x_{1\perp} d^2x_{2\perp} e^{-i(x_{1\perp}\cdot p_{1\perp}+x_{2\perp}\cdot p_{2\perp})}\left( 1 - \frac{1}{N_c}\Tr\left[ U_1U_2^\dagger\right]\right) =\delta^2(p_{1\perp})\delta^2(p_{2\perp}) - \frac{1}{N_c}\Tr\left[ \tilde{U}_1\tilde{U}_2^\dagger\right],
        \end{equation}
         where $\tilde{U}$ is the fundamental representation of the Wilson line in momentum space, labeled by a subscript that corresponds to the respective $t$-channel momentum attributed to it (either $p_1$ or $p_2$). $\Phi_{q\bar{q}}^{\mu}$ is the projectile impact factor which reads
        \begin{equation}
			\label{eq:phi1}
			\Phi^{\mu}_{q\bar{q},1} = - \frac{\alpha_k \left[ \Bar{u}_{p_q} \gamma^+ (p^+_q\gamma^- +\hat{p}_{q1\; \perp}) \gamma^{\mu} \left((p_{\Bar{q}}^+ + k^+)\gamma^- + \hat{p}_{\Bar{q}2\perp} +\hat{k}_\perp\right)\gamma^+ ( \hat{p}_{\Bar{q}} +\hat{k}) \hat{\epsilon}^*_T v_{p_{\Bar{q}}}\right]}{2p_{\gamma}^+\alpha_{\Bar{q}} \alpha_q(\alpha_{\Bar{q}} + \alpha_k) \left( Q^2 + \frac{\Vec{p}_{q1}^2}{\alpha_q(1-\alpha_q)}\right) \left(\Vec{k} - \frac{\alpha_k}{\alpha_{\Bar{q}}} \Vec{p}_{\Bar{q}} \right)^2}\,,
		\end{equation}
        in the case of emission after the shockwave, and
        \begin{align}
				\label{eq:phi2}
				\Phi^{\mu}_{q\bar{q},2}& = \Bar{u}_{p_q} \gamma^+ (p_q^+\gamma^- + \hat{p}_{q1\; \perp}) \gamma^{\mu}\nonumber
				\\
				& \qquad\times\left[\frac{\gamma^-(p^+_{\Bar{q}} +k^+) - \gamma^+ \frac{\alpha_qQ^2 + \Vec{p}^2_{q1}}{2p_q^+} + \hat{p}_{\Bar{q}2\perp} +\hat{k}_{\perp}}{2p_{\gamma}^+ \alpha_q \alpha_{\Bar{q}} (\alpha_{\Bar{q}} + \alpha_k) \left(Q^2+\frac{\Vec{p}_{q1}^2}{\alpha_q} + \frac{\Vec{p}_{\Bar{q}2}^2}{\alpha_{\Bar{q}}} + \frac{\Vec{k}^2}{\alpha_k}\right) \left( Q^2 + \frac{\Vec{p}_{q1}^2}{\alpha_q(1-\alpha_q)}\right)} \right]
				\hat{\epsilon}^*_T (p_{\Bar{q}}^+\gamma^- + \hat{p}_{\Bar{q}2\; \perp}) \gamma^+ v_{p_{\Bar{q}}}\,,
			\end{align}
        in the case of emission before the shockwave. In the above, we have used the shorthand $p_{ab} = p_{a} - p_{b}$ and defined the ratios $\alpha_{q(\bar q)} = \frac{p_{q(\bar q)}^+}{p_\gamma^+}$.

        \subsection{Pion Distribution Amplitude}
        \label{subsec:Pion-Dist-Amp}
        Next, we project the quark-antiquark pair of the projectile impact factor onto a neutral pion final state. This involves factorizing the Dirac and color structure through Fierz identity. The result is the impact factor in terms of a convolution over the momentum fraction $z$ of the pion carried by the quark,
        \begin{equation}
         \label{eq:projection-pi0}
             \Phi^{\mu}_{q\bar{q}} \to \Phi^{\mu}_{\pi^0} = -\frac{i f_{\pi^0}}{4\sqrt{N_c}}\int dz\, \,\phi_{\pi^0}(z) \Tr\left[H^{\mu} \hat{p}_{\pi} \gamma^5 \right]\,,
         \end{equation}
         where $H^{\mu}$ is defined by
          \begin{equation}
        \Phi_{q\bar{q}}^\mu = \bar{u}_{p_q} H^{\mu} v_{p_{\bar{q}}}\,,
    \end{equation}
    and $f_{\pi^0}$ is the neutral pion decay constant.
        To model the pion distribution amplitude $\phi_{\pi^0}$, we use the simple asymptotic form given by
        \begin{align}
        \label{eq:DA}
            \phi_{\pi^0}(z) = 6 z (1-z)\,.
        \end{align}
         The next step is to compute the traces of Dirac gamma matrices. Since we focus on the photoproduction limit, we only show the result when the open index $\mu = i$, with $i$ being a transverse index. The result is
         \begin{align}
				\label{eq:trace-phi1-perp}
				\Tr\left[  H_{1\perp}^i \hat{p}_\pi \gamma^5\right] &=  -4i p_\gamma^+\alpha_k\left[ (zp_{\pi\perp}-p_{1 \perp})_j \left( \frac{\epsilon_{k\perp}^*\cdot p_{\pi\perp}}{\alpha_\pi} + \frac{\epsilon_{k\perp}^*\cdot(\bar{z}p_{\pi\perp}+k_\perp)}{1-z\alpha_\pi} -2\frac{\epsilon_{k\perp}^* \cdot k_{\perp}}{\alpha_k} \right) \varepsilon^{+-ij} \right. \nonumber
				\\ 
				& \qquad\left. + \; (1-2z\alpha_{\pi}) (zp_{\pi\perp}-p_{1\perp})^i \left( \frac{p_{\pi\perp j}}{\alpha_\pi} -\frac{(\bar{z}p_{\pi\perp}+k_\perp)_j}{1-z\alpha_\pi}\right) \varepsilon^{+-j\epsilon_{k\perp}^*} \right] \nonumber
				\\
				& \qquad \times \left( z\bar{z} \alpha_\pi\left(Q^2+ \frac{(z\vec{p}_\pi - \vec{p}_1)^2}{z\alpha_\pi(1-z\alpha_\pi)}\right)\left( \vec{k} -\frac{\alpha_k}{\alpha_\pi}\vec{p}_\pi\right)^2 \right)^{-1}\nonumber
				\\[5pt]
				 \qquad &= 4ip_\gamma^+\mathcal{H}_{1\perp}^i (z,Q^2,p_{1\perp})\,,\\[7pt]
			\label{eq:trace-phi2-perp}
				\mathrm{Tr}\left[H_{2\perp}^i \hat{p}_\pi \gamma^5\right] &= -4ip_{\gamma}^+ \left[ z\bar{z}\alpha_\pi^2 \left(Q^2\epsilon_{k\perp j}^* +\left(\frac{2\epsilon_{k\perp}^*\cdot k_\perp}{\alpha_k} - \frac{\epsilon_{k\perp}^* \cdot (\bar{z}p_{\pi\perp}-p_{2\perp})}{\bar{z}\alpha_\pi}\right)\frac{(z p_{\pi\perp} - p_{1\perp})_j}{z\alpha_\pi} \right)\varepsilon^{+-ij} \right.\nonumber
				\\ 
				& \qquad \left. - (zp_{\pi\perp} - p_{1\perp})^i ( \bar{z}p_{\pi\perp j} - 2\bar{z}\alpha_{\pi}p_{1\perp j} - (1-2z\alpha_{\pi})p_{2\perp j}) \varepsilon^{+-j\epsilon_{k\perp}^*} \right]\nonumber
				\\
				& \qquad \times\left(\frac{1}{ z\bar{z}\alpha_{\pi}(1-z\alpha_\pi)\left( Q^2+ \frac{(z\vec{p}_\pi - \vec{p}_1)^2}{z\alpha_\pi(1-z\alpha_\pi)}\right)\left(Q^2 + \frac{(z\vec{p}_\pi - \vec{p}_1)^2}{z\alpha_\pi} + \frac{(\bar{z}\vec{p}_\pi - \vec{p}_2)^2}{\bar{z}\alpha_\pi} + \frac{\vec{k}^2}{\alpha_{k}}\right)  }\right)\nonumber
				\\[5pt]
				&  =  4ip_\gamma^+\mathcal{H}_{2\perp}^i (z,Q^2,p_{1\perp})\,.
		\end{align}
       The remaining two traces in which the photon is instead emitted from the quark amounts to simply an exchange in $p_{1\perp}\leftrightarrow p_{2\perp}$ and $z\leftrightarrow \bar{z}$. Moreover, there is a reversal in the order of gamma matrices in the trace, which leads to a  minus sign. However, the projection onto the pion DA requires inserting a Dirac structure $\hat{p}_\pi \gamma^5$, see \EQ\eqref{eq:projection-pi0}, which should not be flipped. Taking this into account, we find that the net effect is that there is  no sign difference between diagrams where the quark and anti-quark are swapped. If one further assumes the forward scattering limit in which $p_{1\perp}=-p_{2\perp}$ (${\Delta_\perp} = 0$), one finds that the hard parts  in \EQ\eqref{eq:trace-phi1-perp} and \EQ\eqref{eq:trace-phi2-perp} are identical to their swapped counterparts. Therefore, in the forward limit, the full reduced amplitude is described by the two diagrams alone in \FIG\ref{fig:LO-diag-1}, multiplied by a factor of 2.
        
        \subsection{Constructing the amplitude and cross section}
        \label{subsec: Constructing 
    Amplitude}
  The non-perturbative model for the dense target is determined by the following dipole operator wedged between initial and final states of the target,
\begin{equation}
			\label{eq:target-matrix-element}
			\langle N^\prime|\tilde{\mathcal{U}}_{12}| N \rangle = -\frac{(2\pi)^{3}}{N_c} \delta(p^-_{N^\prime N})\delta^2(p_{N^\prime N\perp}+p_{1\perp} + p_{2\perp}) \tilde{\mathbf{F}}\left(\frac{p_{12\perp}}{2} \right).
		\end{equation}
        After projecting the $q\bar q$ pair onto a pion final state as discussed in \SEC\ref{subsec:Pion-Dist-Amp} (indicated by the right arrow below), and extracting the relevant normalizations and Dirac delta functions, the canonical form of the amplitude for $\gamma N \rightarrow \pi^0\gamma N^{\prime}$ is
        \begin{equation}    \label{eq:FullAmpPionPhoton}
			\langle N^\prime|M_{q\bar{q}}| N \rangle \rightarrow \mathcal{M} = \frac{(2\pi)^4\delta^4(p_\pi+k+p_N-p_{N^\prime} -p_\gamma)}{ \sqrt{2p_\gamma^+}\sqrt{2k^+}\sqrt{2p_q^+} \sqrt{2p_{\bar{q}}^+} \sqrt{2p_N^-} \sqrt{2p_{N^\prime}^-}}\mathcal{T}.
        \end{equation}
        The transverse reduced amplitude is
        \begin{align}
            \mathcal{T}_T = \left(\frac{2p_N^-}{(2\pi)^4}\right)\left(\frac{e_q^2\sqrt{N_c}}{2\pi}\right)\left(-\frac{(2\pi)^3}{N_c}\right)
            \epsilon_{\gamma\perp i}\int d^2p_{1\perp}d^2p_{2\perp}\delta^2(p_{N^\prime N\perp}+p_{1\perp} + p_{2\perp})\tilde{\mathbf{F}}\left(\frac{p_{12\perp}}{2}\right)\Phi_{\pi^0\perp}^i(p_{1\perp},p_{2\perp}).
        \end{align}
        Note, the prefactors in the first set of parentheses come from pulling out $\frac{(2\pi)^4}{\sqrt{2p_N^-}\sqrt{2p_{N^\prime}^-}}$ in \EQ\eqref{eq:FullAmpPionPhoton} and preemptively using the Dirac delta $\delta(p^-_{N^\prime N})$ seen in \EQ\eqref{eq:target-matrix-element}. Making the substitution of the impact factor in \EQ\eqref{eq:projection-pi0}, with the pion DA in \EQ\eqref{eq:DA} and the traces in \EQs\eqref{eq:trace-phi1-perp} and \eqref{eq:trace-phi2-perp}, and upon carrying out the integration over $p_{2\perp}$, we obtain
        \begin{equation}
			\label{eq:full-reduced-amp}
			\mathcal{T}_{T} = \left(-\frac{6 sf_{\pi^0}e_q^2}{(2\pi)^{2}N_c} \right)\epsilon_{\gamma\perp\mu}\int d^2p_{1\perp} \,\tilde{\mathbf{F}}\left(p_{1\perp}\right)
			\int_0^1 dz \;z(1-z)\mathcal{H}^{\mu}_{\perp}(z,Q^2,p_{1\perp}).
		\end{equation}
        Here, we have a convolution of the perturbative hard part ${\cal H}^\mu_{\perp}(z,Q^2,p_{1\perp})$ with a pion distribution amplitude defined in \EQ\eqref{eq:DA}, and a non-perturbative model of the target impact factor ${\tilde{\mathbf{F}}}\left(p_{1\perp}\right)$. 
         The differential cross section in the photon-target frame may then be written as
        \begin{equation}
			d\sigma_{\gamma N \rightarrow\gamma\pi^0 N'} = \frac{(2\pi)^4}{4s}\delta^4(p_\gamma+p_N -p_\pi-k-p_{N^\prime})|\mathcal{T}_T|^2dLIPS \,.
		\end{equation}
        The tensor structure of the amplitude in \EQ\eqref{eq:full-reduced-amp} can be simplified through Schouten identities and Lorentz invariance, as well as by assuming the photoproduction limit ($Q^2=0$) and the forward scattering limit ($\Delta_\perp = 0$). Then, upon explicitly employing the linear polarization vectors in \EQ\eqref{eq:linear-polar}, we find  that there are only two polarization configurations which are non-zero,
        \begin{align}
        \label{eq:AT}
                 \mathcal{T}_{xy} &= \;2\mathcal{C}_T \int d^2p^{\prime}_1 \tilde{\mathbf{F}}\left(\vec{p}^{\prime}_{1}\right) \int_0^1dz \left(\frac{z}{\vec{p}_1^{\,\prime 2}}\right)\nonumber
                \\&\qquad
                \times \left[ (1+\alpha_{\pi}-2z\alpha_{\pi})\left(\frac{1}{\alpha_{\pi}D^{\prime}_1}  -\frac{1}{\alpha_kD^{\prime}_2}\right)\vec{p}_t\cdot\vec{p}_1^{\,\prime} +\frac{2\alpha_{\pi}(\vec{p}_t\cdot\vec{p}^{\,\prime}_1)^2}{\vec{p}_t^{\,2}D^{\prime}_2}+ \frac{(\alpha_k - \alpha_{\pi})\vec{p}_1^{\,\prime 2}}{D^{\prime}_2}\right]\,, \\
               \label{eq:BT}
               \mathcal{T}_{yx} &= \;2\mathcal{C}_T \int d^2 p^{\prime}_1 \tilde{\mathbf{F}}\left(\vec{p}^{\prime}_{1}\right) \int_0^1dz \left(\frac{z}{\vec{p}_1^{\,\prime2}}\right)\nonumber
                \\ &
                \qquad \times \left[-\alpha_k(1-2z\alpha_{\pi})\left(\frac{1}{\alpha_{\pi}D^{\prime}_1} - \frac{1}{\alpha_kD^{\prime}_2}\right)\vec{p}_t\cdot\vec{p}_1^{\,\prime}
                +\frac{2\alpha_{\pi}(\vec{p}_t\cdot\vec{p}^{\,\prime}_1)^2}{\vec{p}_t^{\,2}D^{\prime}_2} - \frac{\vec{p}_1^{\,2}}{D^{\prime}_2}
                \right]\,,
                \end{align}
        where 
        \begin{equation}
            \label{denoms}
                D^{\prime}_1 = \frac{\vec{p}_t^{\,2}}{\alpha_{\pi}^2}\,,\qquad D_2^{\prime} = \frac{\vec{p}_t^{\,2}}{\alpha_{\pi}\alpha_k} +\frac{(\vec{p}^{\,\prime}_1-z\vec{p}_t)^2}{z(1-z)\alpha_{\pi}}\,,
        \end{equation}
         and $\mathcal{C}_T$ is the collection of prefactors
         \begin{equation}
			\mathcal{C}_T = -\frac{6 s f_{\pi^0} \alpha_{em}Q_q^2}{\pi N_c}.
		\end{equation}
        The first (second) index on the ${\cal T}$ above corresponds to the choice of polarization vector of the outgoing (incoming) photon, see \EQ\eqref{eq:polarization-vectors}.
         Note that we have also performed a shift in the $t$-channel momentum, $\vec{p}_1 \rightarrow \vec{p}_1^{\,\prime}-z\vec{p}_t$, in order to avoid spurious divergences during numerical integration. In addition, since we will compare the amplitude for different numerical models constructed in coordinate space, it is convenient to define the amplitude in terms of the inverse Fourier transformed impact factor. For brevity, only one of the polarization configurations is shown,
         \begin{align}
             \mathcal{T}_{xy} &= 2\,\mathcal{C}_T \int d |\vec{r}|\;|\vec{r}| \mathbf{F}(|\vec{r}|)\int_0^1 dz  \left[  z^2(1-z)\alpha_{\pi}K_0\left( |\vec{r}||\vec{p}_t|\sqrt{\frac{z(1-z)}{\alpha_k}} \right) \right.\nonumber
             \\
              & \quad \left. +\frac{(1-z\alpha_{\pi})\alpha_{\pi}}{|\vec{p}_t|^2}\left( \int_0^{z|\vec{p}_t|}d|\vec{p}_1| \frac{|\vec{p}_1|^3J_0(|\vec{r}||\vec{p}_1|)}{|\vec{p}_1|^2+\frac{(1-z)z}{\alpha_k}|\vec{p}_t|^2}\right) - z^2(1-z)\alpha_{\pi}^2\left(\int_{z|\vec{p}_t|}^\infty d|\vec{p}_1| \frac{|\vec{p}_1|J_0(|\vec{r}||\vec{p}_1|)}{|\vec{p}_1|^2+ \frac{(1-z)z}{\alpha_k}|\vec{p}_t|^2}\right) \right].
            \end{align}

        \section{Numerical Analysis}
        \label{Numerical Analysis}
        
         The integration over the $t$-channel momentum $|\vec{p}_1|$ and quark momentum fraction $z$ in \EQ\eqref{eq:AT} and \EQ\eqref{eq:BT} is performed numerically, whereas the integration over the polar angle of $\vec{p}_1$ is performed analytically. There are three independent kinematic variables in which to express the amplitude and cross section; here, we use the center of mass energy in the photon target frame $s_{\gamma N}$, the magnitude of the back-to-back transverse momentum $|\vec{p}_t|$ of the pion-photon system (forward limit, $\vec\Delta = 0$), and the rapidity of the pion $\eta$. 
         
         The following analysis is then two-fold. In \SEC\ref{Amplitude and Vanishing Point}, we discuss the general structure and magnitude of the amplitude. To model the target impact factor, we use the GBW model as a simple phenomenological model that accounts for saturation. It is defined in terms of the dipole cross section 
         through a critical dipole radius
        \begin{equation}
			R_0(x_{\mathbb{P}}) = \frac{1}{Q_s(x_{\mathbb{P}})} = \left( \frac{x_{\mathbb{P}}}{x_0}\right)^{\lambda/2} \frac{1}{Q_0} =  \left(\frac{|\vec{p}_t| e^{-\eta}}{x_0(\sqrt{s}-|\vec{p}_t|e^{\eta})}\right)^{\lambda/2} \frac{1}{1\,\mathrm{GeV}},
		\end{equation}
        which is related to the saturation scale $Q_s$. In the above, we have used $x_{\mathbb{P}}$-Pomeron in the photoproduction limit, given by
         \begin{equation}
			x_{\mathbb{P}}\approx \frac{M_{\pi^0\gamma}^2}{s} = \frac{\vec{p}_t^2}{\alpha_\pi\alpha_ks}.
		\end{equation}
        Also, we have chosen $Q_0 = 1 \,\mathrm{GeV}$ and used the relationship $\alpha_\pi \approx \frac{|\vec{p}_t|}{\sqrt{s}}e^\eta$.
        The target model in momentum space is then defined as
        \begin{equation}
			\tilde{\mathbf{F}}\left(\frac{p_{12\perp}}{2}\right) = (2\pi)^2 N_c\sigma_0\left(\delta^2\left(\frac{p_{12\perp}}{2}\right) -\frac{R_0^2}{\pi}\exp\left[-\frac{\vec{p}_{12}^{\,2}R_0^2}{4} \right]\right) = (2\pi)^2 N_c\sigma_0 \mathcal{F}\left(\frac{p_{12\perp}}{2}\right), 
		\end{equation}
        where common values for the fitted parameters are \cite{Golec-Biernat:1998zce}
        \begin{equation}
			\sigma_0 = 23.02 \,\mathrm{mb}\,, \qquad x_0 = 3.04\times10^{-4}\,, \qquad \lambda = 0.288\,.
		\end{equation}

        Then, in \SEC\ref{Saturation vs Non-saturation}, we discuss the behavior of the amplitude and cross section using different models for the target in order to compare the effects of saturation vs no saturation.

        \subsection{Amplitude and vanishing point}
        \label{Amplitude and Vanishing Point}
               
        \begin{figure}[!h]
        \centering
        \begin{minipage}[b]{1\textwidth}
        \includegraphics[width=\textwidth]{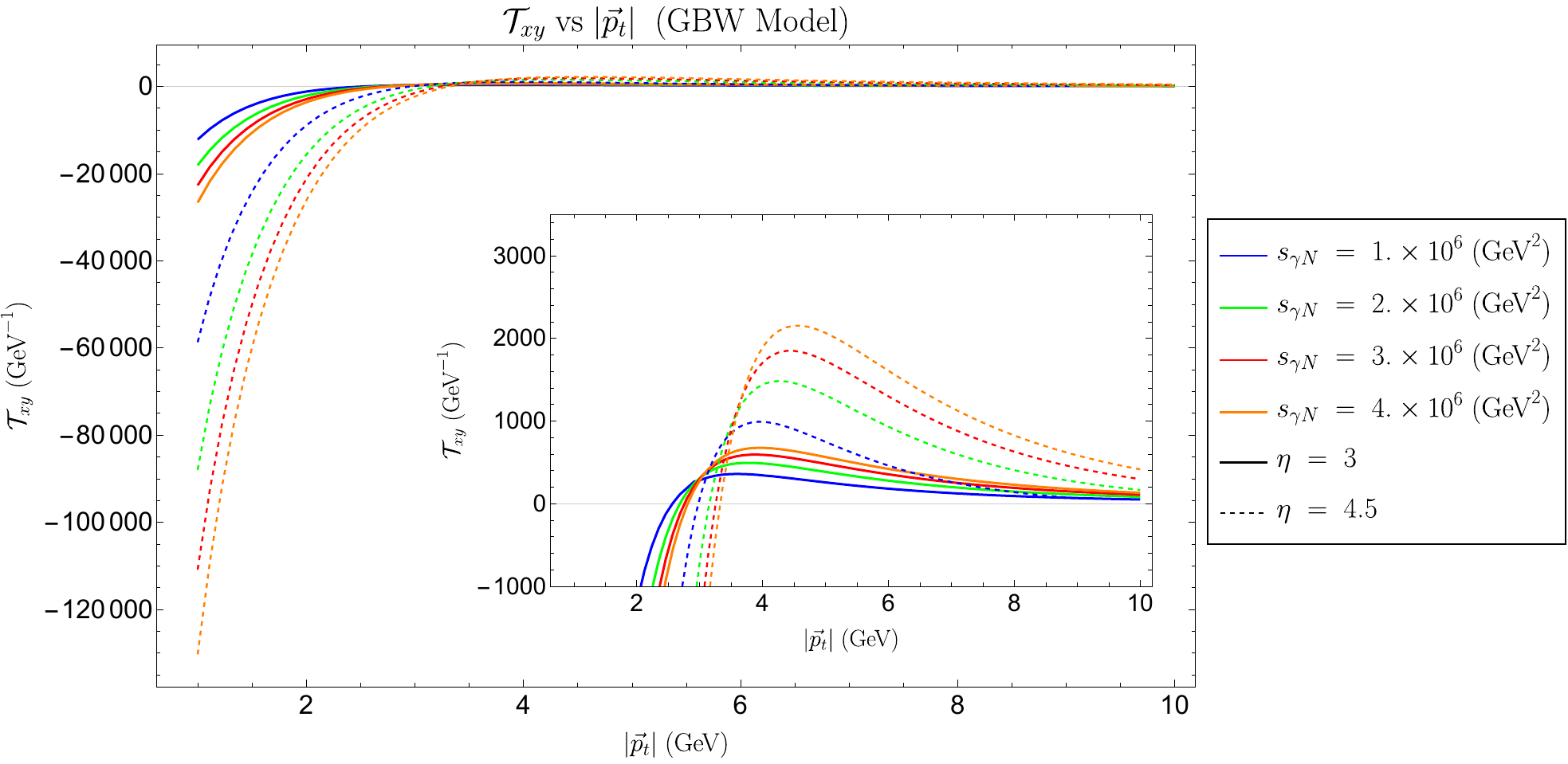}
        \end{minipage}
        \caption{The polarized amplitude ${\cal T}_{xy}$, convoluted with the GBW model for select values of pion rapidity $\eta$ and photon-target center of mass energy $s_{\gamma N}$, as a function of $|\vec{p}_t|$. The inset plot is the same but over a narrow range of the vertical axis to better illustrate the changing of sign. 
        }
        \label{fig:amp and DC}
        \end{figure}

\begin{figure}[!h]
        \centering
        \begin{minipage}[b]{0.96\textwidth}
        \includegraphics[width=\textwidth]{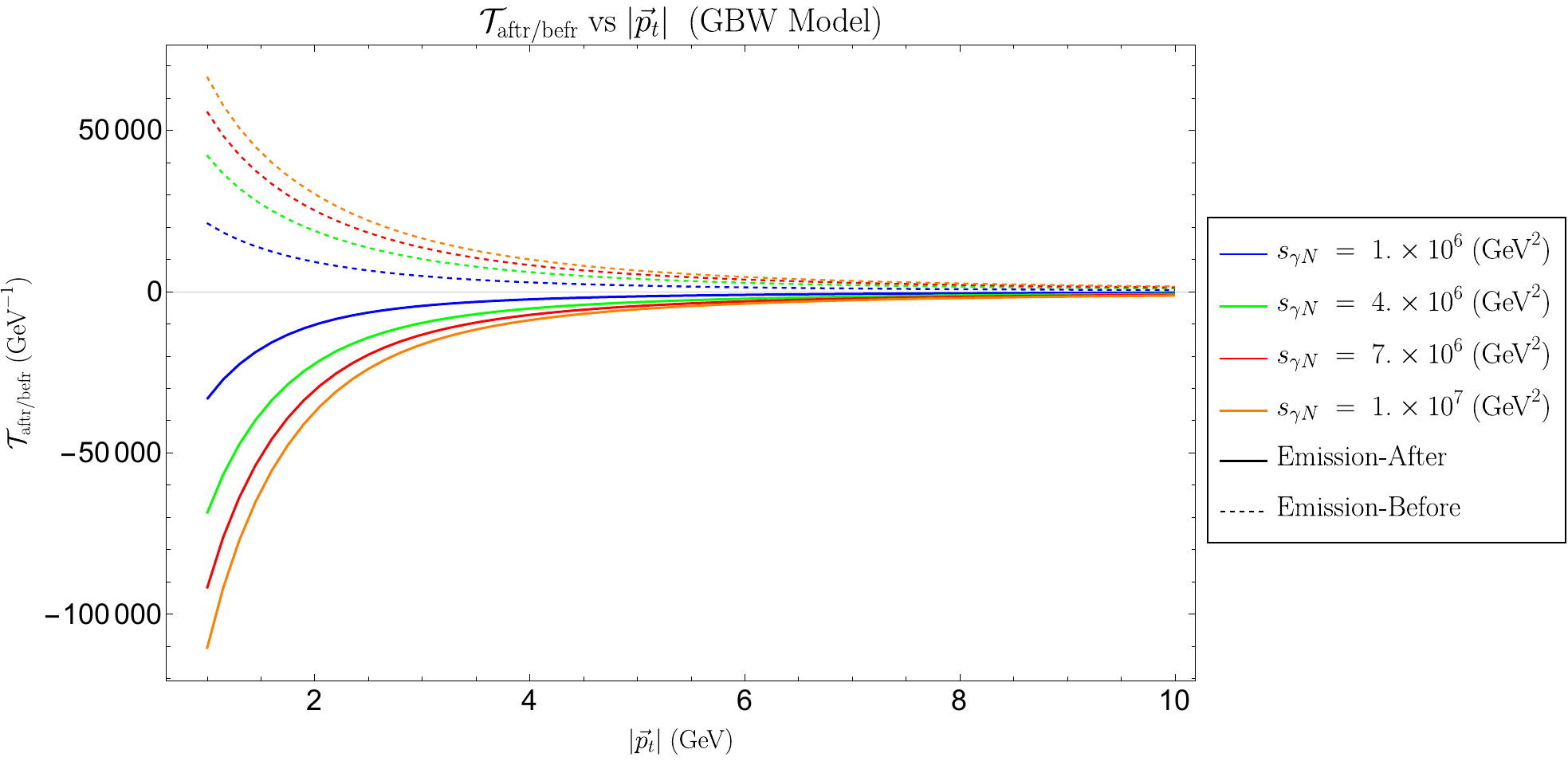}
        \end{minipage}
        \caption{Contributions to the XY polarized amplitude split into those for photon emitted before shockwave (dashed) vs after shockwave (solid), shown as a function of $|\vec{p}_t|$, at fixed pion rapidity $\eta=3$ and for different values of center of mass energy $s_{\gamma N}$.} 
        \label{fig:SeparateDiagrams}
        \end{figure}

         The polarized amplitude in the photon-target frame as a function of $|\vec{p}_t|$ for one of the two polarization configurations is shown in \FIG\ref{fig:amp and DC}. The most notable feature is the change in sign of the amplitude. Since the target impact factor and pion distribution amplitude are both strictly positive, this implies that an exact cancellation between terms has to occur in the projectile impact factor at a specific choice of kinematic variables. Therefore, considering the terms in square brackets of \EQ\eqref{eq:AT}, we break down the individual contributions to the amplitude. Notably, the terms which contain $D_1^\prime$ in the denominator are the contributions associated to the diagram in which the outgoing photon is emitted after the shockwave (\FIG\ref{fig:LO-diag-1} left), whereas terms which contain $D_2^\prime$ are associated to photon emission before the shockwave (\FIG\ref{fig:LO-diag-1} right). These two contributions to the amplitude are shown separately in \FIG\ref{fig:SeparateDiagrams}. We thus find that, in this particular polarization configuration, the ``emission-after'' diagram is always negative while the ``emission-before'' diagram is always positive.

        \subsection{Comparison of results with and without saturation}
        \label{Saturation vs Non-saturation}
        Given the distinct nature of the impact factor shown in \SEC\ref{Amplitude and Vanishing Point}, we now investigate how the amplitude behaves when convoluted with different models for the target impact factor, particularly to investigate the effects of saturation. To do so, we will rely on numerical models whose initial conditions are fit using proton DIS data \cite{Siddikov:2025orq,Benic:2024pqe} and evolved using the BFKL equation for the case of non-saturation and BK equation for saturation.
        
        \begin{figure}[!ht]
        \centering
        \begin{minipage}[b]{0.95\textwidth}
        \includegraphics[width=\textwidth]{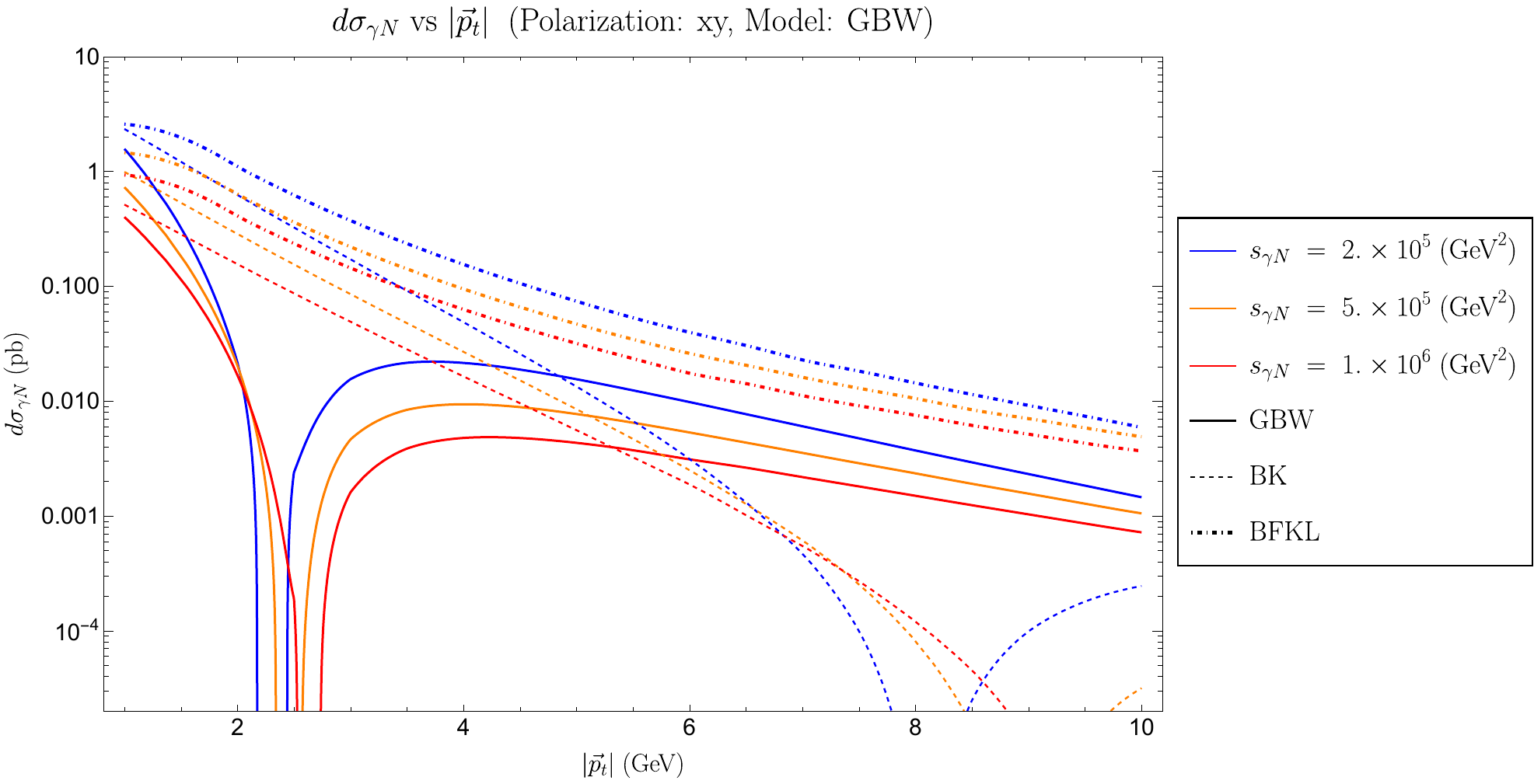}   
        \end{minipage}
        \caption{The differential cross section for XY polarization at fixed pion rapidity $\eta=3$, and for 3 different values of center of mass energy $s_{\gamma N}$. They compare the result of convoluting the projectile impact factor with several models including those which include saturation (GBW and BK) and those that do not (BFKL) in the evolution, see main text.}
        \label{fig:DC Model Compare}
        \end{figure}
        The comparison of the differential cross section through convolution with the three different models for the target impact factor is shown in \FIG\ref{fig:DC Model Compare}. The most notable difference is that the saturation models  have the feature of a vanishing point in the polarized amplitude, whereas the non-saturation model does not. 
         Although this feature is intriguing, it is important to emphasize that the linear (BFKL) predictions were obtained from the BK evolution equation with the non-linear term set to zero, and by assuming the \textit{same} initial condition for both dynamics.
         The nonperturbative parameters of this initial condition were determined by fitting the BK-evolved MV model to the DIS data presented in Ref.~\cite{Lappi:2013zma}. A more definitive assessment of the differences between the linear and nonlinear dynamics will require re-fitting the initial condition to the DIS data using the BFKL evolution equation instead of the BK one. We plan to address this issue in a forthcoming work.

        \section{Conclusion}
        In this work, we studied for the first time the exclusive photoproduction of a $\pi^0 \gamma$ pair. As this process violates collinear factorization at the leading twist, with the breakdown already manifest at leading order, we employed a hybrid framework that combines high-energy and collinear factorization. Within this framework, the diffractive photon-quark–antiquark  production ($\gamma N \rightarrow q \bar{q} \gamma N'$) at small $x$ is treated using the shockwave formalism, while the $q \bar{q} \rightarrow \pi^{0}$ transition is described using collinear factorization, through the pion distribution amplitude. A preliminary phenomenological analysis suggests that this process is promising for exploring small-$x$ dynamics and deserves further study.

\section*{Acknowledgments}

    This work was supported by the GLUODYNAMICS project funded by the ``P2IO LabEx (ANR-10-LABEX-0038)'' and from the "P2I - Graduate School of Physics", in the framework ``Investissements d’Avenir'' (ANR-11-IDEX-0003-01) 
managed by the Agence Nationale de la Recherche (ANR), France. This work was also supported in part by the European Union’s Horizon 2020 research and innovation program under Grant Agreements No. 824093 (Strong2020). This project has also received funding from the French Agence Nationale de la Recherche (ANR) via the grant ANR-20-CE31-0015 (``PrecisOnium'')  and was also partly supported by the French CNRS via the COPIN-IN2P3 bilateral agreement and via the IN2P3 project “QCDFactorisation@NLO”. The work of S. N. was partly supported by the Science and
Technology Facilities Council (STFC) under Grant
No. ST/X00077X/1, and 
by the Royal Society through Grant No. URF/R1/201500. L. S. was supported by the Grant No. 2024/53/B/ST2/00968 of the National Science center in Poland. The work of M.~F. is
supported by the ULAM fellowship program of NAWA No. BNI/ULM/2024/1/00065 “Color glass condensate effective theory beyond the eikonal approximation”.

\section*{References}

\bibliography{masterrefs-old.bib}

\end{document}